\documentstyle[12pt,aasms4]{article}

\tighten

\slugcomment{submitted to ApJ.}
\newcommand{\mpc}{\rm {h^{-1}Mpc }}
\newcommand{\etal}{{\it et al.\ }}

\newcommand{\omav}{\bar{\omega}}

\newcommand{\tP}{\tilde{P}}

\newcommand{\avg}[1]{\langle{#1}\rangle}
\newcommand{\abs}[1]{\mid{#1}\mid}

\newcommand{\ltsima}{$\; \buildrel < \over \sim \;$}
\newcommand{\lsim}{\lower.5ex\hbox{\ltsima}}
\newcommand{\gtsima}{$\; \buildrel > \over \sim \;$}
\newcommand{\gsim}{\lower.5ex\hbox{\gtsima}}

\begin{document}

\title{Higher Order Statistics from the EDSGC Survey I:\\
       Counts in Cells}

\author{Istv\'an Szapudi\altaffilmark{1}}
\affil{Fermi National Accelerator Laboratory}
\affil{Theoretical Astrophysics Group}
\affil{Batavia, IL 60510}

\author{Avery Meiksin\altaffilmark{2} and Robert C. Nichol\altaffilmark{3}}
\affil{University of Chicago}
\affil{Department of Astronomy \& Astrophysics}
\affil{5640 South Ellis Avenue}
\affil{Chicago, IL\ 60637}
\altaffiltext{1}{E-mail:\ szapudi@astro1.fnal.gov}
\altaffiltext{2}{Edwin Hubble Research Scientist; E-mail:\
meiksin@oddjob.uchicago.edu}
\altaffiltext{3}{E-mail:\ nichol@huron.uchicago.edu}

\begin{abstract}
Counts in cells are used to analyse the higher order properties
of the statistics of the EDSGC survey. The probability distribution
is obtained from an equal area projection source catalog
with infinite oversampling over the range $0.015^\circ-2^\circ$.
The factorial moments of the resulting distribution
and the  $s_N$'s characterizing the non-Gaussian nature
of the distribution are extracted. These results are
compared to previous results from the APM survey, and to
theoretical results from perturbation theory. The deprojected
3D values corresponding to the $s_N$'s are also determined. We find
that the 3D values match the scaling relation for strongly nonlinear
clustering found in N-body simulations remarkably well, suggesting
that the galaxies are reliable tracers of the underlying mass distribution.
\end{abstract}

\keywords{large scale structure of the universe --- methods: numerical}

\section{Introduction}
A leading hypothesis for the origin of the large-scale structure of
the distribution of galaxies is that it is a consequence of gravitational
instability in an initially homogeneous medium. The $N-$point correlation
functions provide a set of statistics suited for quantifying the expected
departure from homogeneity of the galaxy distribution under this hypothesis
(\cite{peebles80}). The statistical analyses of recent
 galaxy catalogs has tended
to provide support for this scenario. While the 2--point correlation function
has clearly demonstrated the non-Poisson character of the galaxy distribution,
it is not a unique test of gravitational instability since it is reproduced
by a variety of models for structure formation (\cite{peebles93}).
If gravitational
instability dominates the growth of structure, however, then it is possible to
predict a relation between the higher-order correlation functions and the
2--point function. In particular, if the structure is hierarchical in nature,
as expected in the strongly nonlinear limit, then the $N$--point functions are
symmetrized products of $N-1$ 2--point functions (\cite{peebles80}). In
the limit of weakly nonlinear clustering, analytic forms for the amplitudes
in analogous relations between spatial averages of the correlation functions
have been derived (\cite{jbc93}; \cite{cbs94}; Bernardeau 1994a,b).

Angular catalogs offer two advantages over their redshift analogs for measuring
higher-order correlations: their large size and their insensitivity to redshift
distortions. A disadvantage is that, because they are projections of the
galaxy distribution, simplifying assumptions must be made concerning the
clustering of galaxies to make the extraction of the higher order correlations
practical. Thus the analyses of both types of catalogs are complementary.
Measurements of the higher order
correlation functions in angular catalogs have supported the form predicted by
the hierarchical model. The amplitudes, however, have shown some variance,
depending on the method of analysis and the catalog. Szapudi, Szalay, \&
Bosch\'an (1992) confirmed and refined the estimate of Groth \& Peebles (1977)
for the 3--point function of the Lick counts (\cite{sw67}), although their
estimate of the amplitude of the 4--point function falls somewhat below that
of Fry \& Peebles (1978). Szapudi et al. provide estimates for higher order
functions as
well. Analyses of the IRAS catalogs have provided even stronger support for
the hierarchical model, although the correlations of these infrared-selected
galaxies tend to be somewhat weaker than those of their optical counterparts,
perhaps reflecting a genuine morphology-dependence in the
nature of clustering (\cite{mss,bouchet93}).
Recently, the analysis of higher order functions has been extended to the APM
catalog (Maddox et al. 1990a,b,c) by \cite{gaz94} and Szapudi \etal (1995),
(hereafter \cite{sdes95}), with generally good agreement with the Lick results
of Szapudi, Szalay, \& Bosch\'an (1992), although there are some discrepancies.
These may be due to differences in the scales over which these
functions are averaged, but the differences between the catalogs or the
measurement techniques cannot be precluded as the origin. Systematic variations
in the measured magnitudes will induce artificial correlations, while different
techniques will exhibit differing degrees of sensitivity to the sources of
measurement error (Szapudi \& Colombi 1996, hereafter \cite{sc96}).

In this paper, we present an analysis of the higher order functions in the
EDSGC catalog, an angular catalog covering approximately 1000 square degrees
(\cite{heydon89}; \cite{collins92}).
We employ an efficient method based on factorial moments of cell counts.
The infinite sampling of the catalog (\cite{sz96})
eliminates the measurement errors
arising from the use of a finite number of sampling cells
(\cite{sc96}).

In the next section, we describe the EDSGC catalog, followed by an account
of the measurement technique in \S3. We present the results of the analysis
in \S4, and discuss their relation to previous analyses of other catalogs and
to theoretical expectations in \S5.

\section{The Edinburgh/Durham Southern Galaxy Catalogue}

The Edinburgh/Durham Southern Galaxy Catalogue (EDSGC) is a catalogue
of 1.5 million galaxies covering $\simeq1000$ square degrees centered on
the South Galactic Pole (SGP). The database was constructed from
COSMOS scans (a microdensitometer) of 60 adjacent UK IIIa--J Schmidt
photographic plates and reaches a limiting magnitude of $b_j=20.5$.
The entire catalogue has $<10\%$ stellar contamination and is $\gsim95\%$
complete for galaxies brighter than $b_j=19.5$ (\cite{heydon89}).
The two--point galaxy angular
correlation function measured from the EDSGC has been presented by
Collins, Nichol, \& Lumsden (1992) and Nichol \& Collins (1994).

A rectangular area of the catalog between $\alpha=22^h$, passing through
$0^h$ to $3^h$, and declination $-42 \le\delta\le -23$, was suitable for
our purposes. The original coordinates were converted to physical ones
using an equal area projection:\ $x = (\alpha-\alpha_{\rm min})
\cos\delta,\ y = \delta - \delta_{\rm min}$. This simple
formula is suitable for the small angular scales considered in this
paper. The projection did not affect the
declination range, but to obtain a rectangular area the
physical coordinates corresponding to right ascension
were restricted to values
less than $55^\circ$. This resulted in a sample of $2.9\times10^5$ galaxies,
and a total effective survey area of 1045 square degrees, or $\simeq 997$
square degrees after accounting for the cut-out regions.

Magnitude cuts were determined by practical considerations.
The catalog is complete to about $20.3$ magnitude. We
adopt a limit half a magnitude brighter for our analysis
to be conservative. To permit a direct comparison with results from the
APM survey (Gazta\~ naga 1994),  we used the magnitude cut
$16.98 \le m_{\rm EDS} \le 19.8$. There is an offset in the
magnitude scales of the two catalogs (\cite{nichol92}).
Based on matching the surface densities listed in ~\cite{sdes95}, the
magnitude range we have adopted corresponds approximately to the APM
magnitude range $17 \le m_{\rm APM} \le 20$.

\section{The Method of Analysis}

The calculation of the higher order correlation functions consists of
a sequence of three consecutive steps:
estimation of the probability distribution,
calculation of the factorial moments, and extraction of the
normalized, averaged amplitudes of the $N$-point correlation functions.
We present the relevant definitions and theory below.

Let $P_N$ denote the probability that a cell
contains $N$ galaxies, with implicit dependence on the cell
size $\ell$. The best estimator for $P_N$ from
the catalog is the probability that a randomly thrown
cell {\it in the catalog} contains $N$ galaxies
(excluding edge effects, which are negligible for the scales
in the present study, except perhaps on the largest scales as
a result of the holes cut out around bright stars).
This may either be calculated from the configuration of
the points using a computer algorithm (see \cite{sz96}),
or estimated by actually throwing cells at random,
\begin{equation}
  \tP_N  = \sum_{i=1}^C \delta(N_i = N),
\end{equation}
where $C$ is the number of cells thrown and $N_i$ is the
number of galaxies in cell $i$. It is desirable to use
as many cells as possible, since for large $C$, the errors behave as (SC96)
\begin{equation}
  E^{C,V} = (1-\frac{1}{C})E^{\infty,V}+E^{C,\infty},
\end{equation}
where the $E^{C,V}$ is the total theoretical error (not including
the sytematic errors of the catalog), $E^{\infty,V}$ is the `cosmic'
error associated with the finiteness of the catalog, and
$E^{C,\infty}$ is the error associated with the finite number
of cells used for the estimator. Since $E^{C,\infty}\propto C^{-1}$ (SC96),
the lowest possible error is obtained for $C \longrightarrow\infty$.
We employed such a code on scales up to $2^\circ$.

The factorial moments (see e.g. Szapudi \& Szalay 1993),
may be obtained from the probability distribution using
\begin{equation}
  F_k = \sum P_N (N)_k,
\end{equation}
where $(N)_k =  N(N-1)..(N-k+1)$ is the $k$-th falling factorial of $N$.
The $F_k$'s directly estimate the moments of the underlying continuum
random field which is Poisson sampled by the galaxies.
This is equivalent to the ordinary moments after shot noise subtraction
as can be seen from the relation with ordinary moments
\begin{equation}
   \langle N^m \rangle = \sum_{k=0}^mS(m,k) F_k,
\end{equation}
where $S(m,k)$ are the Stirling numbers of the second kind.
The use of factorial moments simplifies all the expressions,
since sums weighted by the Stirling numbers (shot noise) are eliminated.
For instance, the factorial moments of a Poisson distribution
are $F_k = \avg{N}^k$. These could have been obtained from a
constant probability density $\delta(\epsilon-\avg{N})$, which
is the underlying continuum process.
The ordinary moments of the Poisson distribution, however, will
be more complicated, containing `Poisson noise' from the
previous equation. It is worthwhile to note that
we implicitely assume infinitesimal
Poisson sampling throughout this paper.
It is the most widely accepted assumption, although it cannot account for
certain distributions, such as
ones derived from collisions of hard spheres.

The average of the $N$-point angular correlation functions on a scale $\ell$
is defined by
\begin{equation}
  \omav_N (\ell)=A(\ell)^{-N}\int d^2r_1\ldots d^2r_N \omega_N(r_1,\ldots,r_N),
\end{equation}
where $\omega_N$ is the $N$-point correlation function in the two dimensional
survey, and $A(\ell)$ is the area of a square cell of size $\ell$.
We define $s_N$ in the usual way,
\begin{equation}
   s_N = \frac{\omav_N}{\omav_2^{N-1}}.
\end{equation}
This definition is motivated by the assumed scale invariance of the $N$-point
correlation functions in the strongly nonlinear limit (\cite{bs89}),
\begin{equation}
    \omega_N(\lambda r_1,\ldots,\lambda r_N) = \lambda^{-(\gamma-1)(N-1)}
    \omega_N(r_1,\ldots,r_N), \label{eq:whier}
\end{equation}
where $\gamma$ is the slope of the three-dimensional two-point function. The
coefficients also quantify the deviation from gaussian statistics, like
skewness ($N=3$) and kurtosis ($N=4$). Derivations of the coefficients from
perturbation theory have recently been performed in the weakly nonlinear limit
for three dimensions by Juszkiewicz \etal (1993) and Bernardeau (1994a, b), and
for two dimensions by Bernardeau (1995).

The factorial moments have an especially simple relation to
the $s_N$'s through the recursion relation (Szapudi \& Szalay 1993),
\begin{equation}
  s_k = \frac{F_k\omav_2}{N_c^k}-\frac{1}{k}\sum_{q=1}^{k-1}
        \frac{(k-q)s_{k-q}F_q{k \choose q}}{N_c^q},
\end{equation}
where $N_c = \avg{N} \omav_2$.
Note that although the notation indicates a projected catalog,
there are corresponding expressions for three dimensions.

The deprojection of the $s_N$'s to their 3D counterparts has
an intrinsic limitation due to the finite sizes of the cells.
While the deprojection of any individual tree-structure is
well-defined, care must be taken in interpreting the deprojected
values of the cell-count determined $s_N$'s, since these implicitly
contain an averaging over trees within each cell
(see \cite{sdes95} for a discussion). On small scales, where clustering
is strongly nonlinear, the coefficients deproject to the 3D coefficients $S_N$
defined by $S_N=\bar\xi_N/\bar\xi_2^{N-1}$, where the hierarchical
assumption may be presumed valid. In this case,
\begin{equation}
   s_N = R_N S_N, \label{eq:SN}
\end{equation}
where $S_N$ is the corresponding three dimensional value for the spherically
averaged $\bar\xi_N$'s. The projection coefficients $R_N$'s are fairly
insensitive to slight variations of the magnitude cut (see Table 2 in
\cite{sdes95}), and the shape dependence is neglected according to the findings
of Bosch\'an, Szapudi, \& Szalay (1994). We adopt the $R_N$'s of \cite{sdes95}
with a Hubble constant of $H_0=100{\rm km\, s^{-1}\, Mpc^{-1}}$.
In the intermediate range of weakly nonlinear clustering, hierarchy-breaking
terms become significant, and the differences between the conical averaging of
the projected correlation functions and the spherical averaging of the three
dimensional functions become large (\cite{bern95}).
In this limit, the $s_N$ deproject according to
\begin{equation}
   s_N = R_N \Sigma_N, \label{eq:SigN}
\end{equation}
where the $\Sigma_N$'s involve averages only over the orthogonal parts of the
wavevectors. (The expressions for $R_N$ are identical in
equations~[\ref{eq:SN}] and [\ref{eq:SigN}] for power-law power spectra.)
Expressions for $\Sigma_N$ for pure power-law power spectra have been worked
out by Bernardeau (1995). For the depth of the EDSGC,
the weakly nonlinear region corresponds to separations of $\theta\gsim1^\circ$
(see \S4).
\section{Results}

We measured counts in cells by calculating the results corresponding
to an infinite number of square cells, placed according to the
algorithm of \cite{sz96}, with sizes in
the range $0.015125^\circ-2^\circ$
(corresponding to $0.1-13\mpc$ with $D \simeq 370\mpc$,
the approximate depth of the catalog). The largest scale is
limited by the geometry induced by the cutout holes:
the number of available cells would be severely limited
for a measurement on significantly larger scales,
since cells intersecting with the cutout holes were
rejected. The smallest
scale approaches that of galaxy halos for the typical
depth of the catalog. Note that even at the smallest scale,
where the average count is only $0.0645$ per cell, the $s_N$'s
are measured to high accuracy because of the infinite oversampling
and the efficient Poisson subtraction through the use of factorial moments.
By comparison, the practice common in the literature is to stop
at scales four times that at which the Poisson noise starts to dominate,
i.e., where the average count approaches unity. Note that this
method extracts almost all the information available through
cell counts,
except that we did not sample different
orientations of the cells, which in principle could have a slight effect.
However, studying different orientations properly would most likely
involve cutting off more of the existing catalog to prevent
potential weighting problems. In practice this could even
diminish the available information by enhancing cosmic
errors; the thorough examination
of this effect is left for future work.

The results of both infinite and low sampling measurements for
$P_N$ are displayed in Figure 1. The low sampling corresponds
to covering the area with cells once only, i.e. the number of
sampling cells is $C_V = V/v$, where V is the volume of the survey
and $v$ is the volume of the sampling cell at the given scale.
As proved in (\cite{sc96}), the `number of statistically independent
cells' , $C^*$, depends strongly both on scales
and on the aims of the measurement, but for higher order
statistics it is generally much larger than $C_V$.
Therefore to minimize the errors as much as possible
we used infinite oversampling for all measurements in this
paper. A comparison of the
two curves shows the substantial improvement in accuracy achieved
through oversampling.
Note that covering the area fully with $C_V$ number of cells
is not fully equivalent to throwing the same number of random
cells, because these might overlap, thus more effectively sampling
clusters, and in principle decreasing the bias toward low values
visible in Figure 2.

Figure 2 shows the scale dependence of the $s_N$'s determined
from the counts in cells. The solid line corresponds to the
measurements of the entire survey area with high oversampling.
The dotted line is the same measurement with undersampling.
For the error determination we divided the survey into four
equal parts, similar to the approach of \cite{gaz94};
this procedure can overestimate the cosmic error,
because subcatalogs have smaller area, but it could also
lead to underestimation, because the subcatalogs are
not independent volumes (\cite{sc96}).
The squares show the mean of the measurements taken in the
following way: to avoid
the bias introduced by the fact that the $s_N$'s are nonlinear
functions of the factorial moments, the mean of the
moments was taken first, and then the cumulants were calculated .
The error bars, estimated by a determination of the dispersion
of the (possibly biased) $s_N$'s calculated from the factorial
moments from each zone, are shown only for those points for which there was
sufficient physically valid data permitting a determination.
Note that, as mentioned above, the results of this
crude estimate must be taken with caution,
because at certain scales it can either under or overestimate
the true error bars, and the error distribution is non-Gaussian
(\cite{sc96}).
On large scales the squares deviate from the solid line: this is
presumably a result of edge effects. For $s_3$ and $s_4$, the errors range over
$8-36\%$ and $19-56\%$, respectively.
These may be compared with
theoretical estimates for the errors. We base the estimates on the errors of
the correlator moments over the entire catalog, according to \cite{sc96}. For
the first four moments, respectively, the errors are, ranging from large scales
to small, $3-2$\%,\ $7-12$\%,\ $13-45$\%,\ and $23-63$\%. Although there is no
simple formula relating the errors of the moments to the errors of the $s_N$'s,
it is likely that the errors at each order are dominated by the largest
error; i.e., the highest moment. Thus, unless some cancelation
effects are present, the last two values should well represent
the errors on $s_3$ and $s_4$. These compare well with the empirical errors
from the dispersion above.
Possible systematics were also checked for: perturbing
the magnitudes of the galaxies with the measured magnitude zero
point distribution yields virtually identical results.

Figure 2 exhibits two plateaus, one at small scales ($<0.03^\circ$)
and a second at large ($>0.5^\circ$). The large scale plateau is approaching
the width of the survey, and so may merely reflect edge effects. The plateau at
small separations, however, may indicate that the strongly nonlinear
clustering limit has been reached, in which case the hierarchical form
for the angular correlations should apply, for which the coefficients appear
to converge. The values of $s_N$ are provided in Table 1, and the ratios
$s_N/ R_N$ in Table 2.

In order to probe more deeply into the weakly nonlinear regime,
we performed a separate analysis extending to $4^\circ$.
On these scales the majority of cells overlaps with
some of the cut-out regions, therefore the analysis had to
be done without the elimination of such cells, otherwise
edge effects and cosmic errors from the resulting small
area would have severly affected the measurement.
After reanalysing all scales without eliminating cells containing
the cut out holes, we found the effect of the holes is to bias
the measurements to slightly low values (dashed line in Figure 2),
but by an amount which is well within the statistical errors.
We nonetheless find good agreement with the smaller
scale analysis for $\theta\leq1^\circ$. We obtain in the larger scale
analysis $s_3=5.75$ and $s_4=60$ at $\theta=2^\circ$, and $s_3=7.9$ and
$s_4=71$ at $\theta=4^\circ$. These correspond to $\Sigma_3=4.95$ and
$\Sigma_4=42$ at $13h^{-1}$ Mpc separation, and $\Sigma_3=6.8$ and
$\Sigma_4=50$ at $26h^{-1}$ Mpc separation (although these
angular scales are outside of the range of strict applicability of the
theory for the $\Sigma_N$'s, ~\cite{bern95}). The errors on these
measurements could be as much as $30\%$ and $50\%$ respectively.

\section{Discussion}

\subsection{Comparison with the APM Catalog}

Figure 3 compares our results for $s_3-s_6$ with estimates
from the APM catalog kindly provided by E. Gazta\~naga.
The heavy solid line extending to the smallest scales is our
measurement for the EDSGC catalog, the dotted lines are
the measurements for the full APM catalog, and the light
solid line is the measurement of a subregion of the
APM which corresponds to the EDSGC. Between scales  of
about $0.2^\circ$ to $2^\circ$ the agreement is good beween the
full EDSGC and the same region of the APM. It seems that
the increase of the $S_N$'s at the largest scales is due to
edge effects: a similar phenomenon appears in the full APM at
larger scales. This figure
again shows that the error bars (which are obtaind from the
APM using a similar zoning procedure), do not necessarily
reflect the true errors: the EDSGC zone of the APM lies
about two sigma outside of the full APM measurements at
the same scales. As was mentioned above, the reason for this
could be both that the estimation of the dispersion by
dividing into four subcatalogs is not accurate, and that
the errors are probably distributed in a non-Gaussian
fashion. At scales smaller than 0.2 degrees,
the APM measurements seem to be systematically low.
The reason for this could be
insufficient sampling in the APM estimates: since the
APM measurement was derived from a density map at the lowest
scale shown at the figure (Gazta\~naga 1996, private
communication), only minimal sampling, and $4$ times
oversampling was used for the two leftmost points,
because this is the most possible with shifting the
grid at these scales. As shown in \cite{sc96} it is
most important to have high oversampling at small
scales, therefore low sampling  could have introduced measurement errors.
Insufficient sampling in principle does not cause a low bias, i.e.
the mean is always recovered as an average over many low
sampling measurements.
However, if the error distribution is
skewed, a realization is likely to undershoot the mean,
which is compensated by a few larger overshoots in
the ensemble average sense.
Physically this corresponds to the fact that
with low sampling it is not likely
to have cells which fully cover small, dense clusters. Since
these clusters dominate the higher order statistics,
insufficient sampling usually results in an underestimate,
and only rarely in an overestimate. This is a possible
explanation both for the effect shown in Figure 2
(dotted line), and for the systematic deviation
between the APM and the EDSGC in Figure 3.

\subsection{Comparison with Theory}

At small (non-linear) scales, the hierarchical tree model
(described below) is believed to be a good approximation
to the clustering.
At larger (weakly non-linear) scales, perturbation theory
of gravitating matter starting from Gaussian initial conditions
provides a basis for comparison. In a projected catalog, the
transition between scales is somewhat uncertain, since a
lengthscale is assigned to angular scales using the depth of the catalog.
This procedure is physically correct although somewhat
arbitrary, and there could be
effects associated with mixing of different scales in the
selection function weighted cone corresponding to a cell
in an angular catalog. While no existing measurement has clearly
demonstrated the validity of either
of the above models, the results based on moment correlators
seem to support the hierarchical model, at least on small
scales (\cite{ssb,mss,sdes95}), as does the present work.
In what follows,
a direct comparison is made without taking into account
the possibility of biasing: the data are consistent
with the galaxies' acting as faithful tracers of the underlying
mass distribution.

A plateau at small separations may be expected when the clustering becomes
strongly nonlinear. The effect is found, for example, in the N-body experiments
for scale-free clustering by Colombi, Bouchet, \& Hernquist (1995). If the
clustering we measure is strongly nonlinear on the smallest scales, then we are
permitted to identify $S_N=s_N/R_N$ in Table 2 at small separations. We
may then in turn derive the 3D clustering amplitudes $Q_N$. These are
defined within the hierarchical model
\begin{equation}
   \xi_N({r_1},\ldots, {r_N}) = \sum_{k=1}^{K(N)}
   Q_{Nk} \sum^{B_{Nk}} \prod ^{N-1} \xi(r_{ij}),
   \label{hierar}
\end{equation}
where $\xi(r) \equiv \xi_2(r) = (r/r_0)^\gamma$, as the average of the $Q_{Nk}$
\begin{equation}
Q_N = \frac{ \displaystyle {\sum_{k=1}^{K(N)} Q_{Nk}B_{Nk} F_{Nk}}}
          { \displaystyle {N^{(N-2)}}},
\end{equation}
where $F_{Nk}$ are the form factors associated with the shape
of cell of size unity (see~\cite{bss94} for details)
\begin{equation}
           F_{Nk} = \int_1 d^3 q_1 \ldots d^3 q_N \prod^{N-1}
        \left\{\abs{q_i- q_j}^\gamma \int_1 d^3 p_1 d^3 p_2 |p_1-p_2|^{-\gamma}
	\right\}^{-1}.
\end{equation}
The product above runs over the $N-1$ edges of a tree.
The summation in equation~(\ref{hierar}) is over
all possible $N^{N-2}$ trees with $N$ vertices.
In the sum, every $\xi(r_{ij})$ corresponds to an edge
$r_{ij}=\abs{{r_i-r_j}}$ in a tree spanned by ${r_1}, \ldots,
{r_N}$. For each tree, there is a product of $N-1$ two-point
functions and a summation over all the $B_{Nk}$ labelings of
all the $K(N)$ distinct trees.

Using the values for $r=0.1$ Mpc in Table 2,
we find for $N=3-9$,
$Q_N=(2.02, 7.3,  30, 108,  320,   745,   1298)$.
The values for $Q_3$ and $Q_4$ somewhat exceed those found for the
Lick-Zwicky sample (\cite{gp77}; \cite{fp78}; \cite{ssb}), and greatly
exceed the values found for the CfA1 and SSRS surveys (\cite{gaz92}). The
discrepancy between the larger angular samples and the smaller samples used
for the redshift surveys has been previously noted by \cite{fg94}. Our results
suggest the discrepancy at small scales may be even larger. The reason for the
difference in the values is unknown, but may be a result of cosmic variance.
It appears not to be a result of the added redshift information, since
Gazta\~naga (1994) found that suppressing the redshifts in the CfA1 and SSRS
surveys and treating the samples as angular catalogs had little effect on
the values.

In the limit of weakly nonlinear clustering, it is possible to compare
the clustering coefficients with theoretical predictions for a given
power spectrum (\cite{jbc93}; Bernardeau 1994a,b; \cite{bern95}).
If $n_{\rm eff}$ is
the local slope of a hypothetical power spectrum that would
yield the measured moments in the weakly non-linear regime,
we find from the values of
$\Sigma_3$ and $\Sigma_4$ at separations of 6.5, 13, and 26 $h^{-1}$ Mpc the
values $n_{\rm eff}=(-1.2, -1.9, -3.1)$ for $N=3$, and $n_{\rm eff}=(-1.3,
-1.7, -1.9)$ for $N=4$, using the results of the larger $4^\circ$ analysis for
$\theta>1^\circ$ from the previous section, and the expressions relating
$\Sigma_N$ to $n$ in Bernardeau (1995) in the small angle
approximation, which is accurate up to scales $1^\circ$. Since
$1^\circ$ corresponds to roughly $6.5$ Mpc, clustering is just entering
the weakly non-linear regime, for which theory and measurement may be
best compared. The trend of decreasing
$n_{\rm eff}$ with increasing scale is suspect. For a power spectrum like
CDM, $n_{\rm eff}$ increases with increasing scale. The inverse trend
may indicate that edge effects are significant on these scales and are
compromising the determination of $s_N$ on scales exceeding $1^\circ$,
or that the theory for $\Sigma_N$ indeed starts to break down.

While the weakly nonlinear limit should break down on scales smaller than
$1^\circ$, it is informative to explore the inferred dependence
of $n_{\rm eff}$ on scale to smaller values. Colombi et al. (1995) find from
N-body experiments for scale-free initial conditions that the values for $S_N$
vary only slightly with scale, increasing for small separations where the
clustering becomes strongly nonlinear. They find, independent of $n$,
\begin{equation}
S_N\simeq\left[D(\bar\xi_2)\right]^{N-2}\tilde S_N, \label{eq:Ssc}
\end{equation}
for $N=3,$ 4, and 5, where $D(\bar\xi_2)=(\bar\xi_2/100)^{0.045}$ and $\tilde
S_N$ is the value of $S_N$ for $\bar\xi_2=100$. The relation implies
a weak departure from the hierarchical clustering behavior, since the $S_N$
depend on scale. The dependence is so weak, however, that the departure is
slight. We compare the clustering amplitudes
found in the EDSGC with this relation in Figure 4a.
 The agreement in the
strong clustering limit is remarkably good, particularly for $N=3$ and 4.
Because we have only angular information, it is not possible to determine
whether the deviation from the scaling relation for $\log\bar\xi_2<2$ is
a real effect or a consequence of the inherent limitations of extracting
3D information from a projected catalog. It should be noted that the agreement
is particularly remarkable since hierarchical clustering is assumed for the
underlying distribution in order to convert from the projected correlations
to the 3D, while the relation of equation~(\ref{eq:Ssc}) violates this
assumption. This suggests that the hierarchical model is a good, though perhaps
not perfect, description of the clustering.

Colombi et al. find that the clustering for $N=3$, 4, and 5 may be
described by a single effective spectral index $n_{\rm eff}$, found from the
expressions for weakly nonlinear clustering (\cite{jbc93}; Bernardeau 1994a,b).
Although the relations between the $S_N$ and $n$ from weakly nonlinear theory
do not apply for strong clustering, and even less so in an angular catalog,
we may adopt them to obtain a formal value for $n_{\rm eff}$ as done by Colombi
et al. (1995, 1996). We do so by fitting $s_N/R_N$ to the expressions for
$S_N$ in the limit of weakly nonlinear clustering for $N = 3\ldots6$ using
least squares, for $\theta\leq1^\circ$. The results are shown in
Figure 4b, including the values derived for each $N$ individually. Within the
error estimates, a single value of $n_{\rm eff}$ appears to provide an adequate
description of the clustering amplitudes, although the errors are large for
small separations. A comparison with N-body results for scale-dependent
clustering models, like a CDM-dominated cosmology, could be very illuminating.

\acknowledgments

I.S. thanks S. Colombi, J. Frieman, and A. Szalay for stimulating discussions.
S. Colombi provided the theoretical error estimates in \S 4.
We are indebted to C. Collins, S. Lumsden, N. Heydon-Dumbleton,
and H. MacGillivray for full use of the EDSGC.
The authors would like to thank the referee, E. Gazta\~naga
for providing his estimates from the APM catalog for comparison,
and for useful suggestions, and E. Wright for suggested improvements.
Additionally, we would like to acknowledge useful discussions with
G. Dalton, J. Loveday, and S. Maddox about properties of the APM
survey.
I.S. was supported by DOE and NASA through grant
NAG-5-2788 at Fermilab. A.M. is
grateful to the W. Gaertner Fund at the University of Chicago for support.

\section{Figure Captions}

\noindent Figure 1.\ Shows the distributions $P_N$ of counts in cells
measured in the EDSGC catalog. The solid line corresponds to infinite
sampling, while the dotted line to severe undersampling.
The curves from left to right correspond to cell sizes
from $0.015125^\circ$ doubling up to $2^\circ$.
Exhaustive sampling is essential on all scales.

\noindent Figure 2.\ The solid line is the measurement of the $s_N$'s
over the entire survey area with infinite sampling, the dotted line
is the same with low sampling. Undersampling results in a
systematic underestimate of the coefficients. The squares show the
mean of the measurements (see text for details) in four equal parts of the
survey, and the errors are calculated from the dispersion.
The misalignment of the squares and the solid line at the largest scales
may be a result of edge effects.
The triangles show the $s_N$'s corresponding
to the best fitting formal $n_{\rm eff}$ (see text).

\noindent Figure 3.\ A comparison with the results from the APM survey
for $s_3 \ldots s_6$. The heavy solid line
is the measurement in the full EDSGC survey, as in Figure 2.
The dotted line is the full APM measurement by Gazta\~naga 1994,
while the light solid line is the measurement in the APM
catalog by E. Gazta\~naga (1996) for the region on the sky
corresponding to the EDSGC.
There is excellent agreement except for the smallest scales,
possibly caused by insufficient sampling in the APM measurements.

\noindent Figure 4.\ a.\ The clustering amplitudes $s_N/ R_N$ as a function
of the average 2-point function $\bar\xi_2$. Also shown is the scaling relation
of Colombi et al. (1995) found in the strongly nonlinear limit in N-body
experiments with scale-free initial conditions.\ b.\ The best formal fits for
$n_{\rm eff}$ ({\it solid}), using up to sixth order quantities.
Also shown are the values determined from each $N$ separately, including an
indication of the errors based on the upper and lower quartile values for
each $S_N$. Within the errors, the clustering may be described by a single
value of $n_{\rm eff}$. Shown are the values of $n_{\rm eff}$ for $N=3$
({\rm dotted}), $N=4$ ({\it short-dashed}), $N=5$ ({\it long-dashed}),
and $N=6$ ({\it dot-dashed}).

\end{document}